\documentstyle[12pt]{article}
\newcommand{\be}{\begin{equation}}
\newcommand{\ee}{\end{equation}}
\newcommand{\bea}{\begin{eqnarray}}
\newcommand{\eea}{\end{eqnarray}}
\newcommand{\ba}{\begin{array}}
\newcommand{\ea}{\end{array}}
\newcommand{\nn}{\nonumber}
\newcommand{\del}{\partial}

\newcommand{\ap}{\alpha^\prime}

\newcommand{\D}{\delta}

\newcommand{\E}{\epsilon}

\newcommand{\s}{\sigma}

\newcommand{\z}{\zeta}

\newcommand{\neu}{{\cal N}}

\newcommand{\vm}[1]{\hspace{#1mm}}
\newcommand{\vc}[1]{\hspace{#1cm}}
\newcommand{\mb}[1]{\mbox{#1}}

\begin{document}
\begin{titlepage}
\begin{flushright}
April, 2002
\end{flushright}
\vspace{0.5cm}
\begin{center}
{\Large \bf 
Partition Function on Not-flat Brane
\par}
\lineskip .75em
\vskip2.5cm
{\large Kayoko FUJIMURA}
\vskip 1.5em
{\large\it Institute of Physics, University of Tsukuba \\
\vspace{1mm}
Ibaraki 305-8571, Japan}
\end{center}
\vskip3cm
\begin{abstract}
 We show that a partition function on the not-flat D1-brane can be
written in the same form as that on the flat one in $\ap$-order.
In this case
the information of the curvature of the brane configuration  
is included in tachyon beta function.
\end{abstract}
\end{titlepage}
\baselineskip=0.7cm

\section{Introduction and Summary}
\vm{4}
On a flat D-brane in the bosonic string theory, 
Born-Infeld action can be obtained as the partition function of the 
two-dimensional nonlinear sigma model action on an open string world sheet.
It is calculated by the path-integral.
Then, the factor like Born-Infeld action
appears in the scattering amplitudes
calculated by the path-integral formalism \cite{hk} \cite{kk}.
In these papers, the scattering amplitudes of closed string states with 
a D0, 1-brane are studied. 
It is shown that the Born-Infeld-like factor 
on the arbitrary configuration of a D-brane
can be written as that in the case of a flat brane 
by renormalizing the embedding function $f^\mu$.

Here we introduce the open string tachyon 
and calculate the partition function 
on the arbitrary configuration of a D-brane.
It is shown that the partition function can be expressed 
as the same factor as 
that on a flat brane up to an $\alpha^{\prime 2}$ order term.
In such a case, however,
the tachyon beta-function is shifted from that in the flat case.    

In this letter
we introduce a open string tachyon field and
try to cancel a contribution from the curvature of the brane configuration.
Following the method of \cite{kk},
we setup tools needed for calculation in Sect. 2.
In Sect. 3
the path-integral is performed
and then we obtain the partition function    
and renormalize a tachyon field.

\section{Setup - Geodesic Normal Coordinate Expansion}

\renewcommand{\theequation}{2.\arabic{equation}}\setcounter{equation}{0}
\vm{4}
We deal with a D1-brane in flat 26-dimensional space-time 
of the bosonic string theory.  
The bosonic open string world-sheet action with a gauge field $A_\mu$ 
and an open string tachyon field $T$ on the boundary is given by \cite{t1}
\bea
 S =  \frac{1}{4\pi\alpha^\prime}\int_{\Sigma}d^2 \bar{\s}
     \del_{\bar{a}} X^{\mu} \del_{\bar{a}} X^{\nu} \eta_{\mu \nu}
  	+ i\int_{\del \Sigma} d\theta A_\mu (X) \del_\theta X^\mu
	+ \frac{1}{\E}\int_{\del \Sigma} d\theta T(X). 
\label{a0}
\eea
In (\ref{a0}) we take the world sheet to be a unit disk $\Sigma$ 
and represent its boundary by $\del \Sigma$,
$\theta$ is the one-dimensional coordinate on $\del \Sigma$,
the subscript $\bar{a}$ of $ \del_{\bar{a}} X^\nu $
 means the coordinate on the world sheet $\bar{\s}$,
$X^\mu$ are the coordinates of the target $26$-dimensional space, 
the metric of the target space is taken as
$\eta_{\mu \nu} = \mb{diag} (-1,1,\cdots 1) $,
and 
$\E$ stands for the short distance cutoff parameter.

Following \cite{hk} and \cite{kk},
 the partition function for the brane configuration $f^\mu (X (\theta) )$ is
evaluated as
\bea
 Z = c \int DX^\mu Dt(\theta) D\sigma(\theta) 
	\delta (X^\mu (\theta) - f^\mu(t(\theta),\sigma(\theta)))
	\exp (-S), 
\label{Z}
\eea
where
$t$ and $\s$ are the coordinates on the D1-brane 
and 
$ c$ is an overall factor independent of $A$ and $T$.

Using the method in \cite{ft1},
we reduce the two-dimensional action to the one-dimensional action on boundary of the world-sheet
\footnote{
It is imposed that $\del F$ and $\del \del \del T$ can be ignored.}:
\bea
 S = \frac{1}{4 \pi \ap } X^\mu \cdot \neu^{-1}_{\mu \nu} \cdot X^\nu
+
i \int_{\del \Sigma} d\theta A_\mu \cdot \del_\theta X^\mu 
+
\int_{\del \Sigma} d\theta \frac{T}{\E} .
\label{one-dim action}
\eea
The Neumann function $\neu$ and its inverse are given by
\bea
\neu^{\mu \nu} = \eta^{\mu \nu} \frac{1}{\pi}
\mathop{ \sum_{k=1}}\frac{1}{k}\mb{e}^{-k\E}\cos k(\theta - \theta^\prime) 
\label{nold} \\
\neu^{-1}_{\mu \nu} = 
\eta_{\mu \nu} \frac{1}{\pi} 
\mathop{ \sum_{k=1}}\mb{e}^{-k\E}k\cos k(\theta - \theta^\prime).
\label{niold}
\eea
The dot in (\ref{one-dim action}) means the following operation;
\bea
A\cdot B (\theta, \theta^\prime )
= \int^{2\pi}_0 d\phi A(\theta,\phi) B (\phi,\theta^\prime). 
\eea

In order to calculate the partition function(\ref{Z}),
we expand $f$ in terms of the geodesic normal coordinates $\z^a$ 
\cite{hk}\cite{kk}\cite{GFM};
\bea
 f^\mu(t(\theta),\sigma(\theta)) = f^\mu(t,\s) 
	& + & \del_a f^\mu (t,\s) \z^a (\theta) 
	+ \frac{1}{2}K^\mu_{ab}(t,\s) \z^a (\theta) \z^b (\theta) \nn \\ 
	& + & 
\frac{1}{3!}K^\mu_{abc}(t,\s) \z^a(\theta) \z^b (\theta) \z^c (\theta)
	+ \cdots,
\eea
where
\bea
   h_{ab} & = &\del_a f^{\mu} \del_b f^{\nu} \eta_{\mu \nu}, \\
   h^{ab} & = & (h^{-1}), \\
 K^{\mu}_{ab} & = & P^{\mu \nu} \del_a \del_b f_{\nu}  \nn \\
             & = & \del_a \del_b f^{\mu} - \del_c f^{\mu} \Gamma^c_{ab} \nn \\
  & = & \nabla_a \nabla_b f^\mu, \\
 \eta^{\mu \nu} & = & h^{\mu \nu} + P^{\mu \nu}, \\
h^{\mu \nu} & = & \del_a f^{\mu} h^{ab} \del_b f^{\nu}.
\eea  
$\nabla$ denotes the covariant derivative for the subscripts a and b,
$ h_{ab} $ the induced metric on the D1-brane,
$\Gamma^c_{ab}$ the Christoffel symbol for the induced metric $h_{ab}$,
$ K^{\mu}_{ab}$ the extrinsic curvature,
and
$P^{\mu \nu}$ and $ h^{\mu \nu}$ stand for projectors
to orthogonal and tangent directions to the D1-brane, respectively. 

Following the background field method, 
we separate $X(\theta)$ into the zero-mode $x$ 
 and the fluctuation $\xi(\theta)$,
\bea
 X^\mu(\theta) = x^\mu + \xi^\mu (\theta) \nn
\eea
with the condition 
\bea
 \oint_{\del \Sigma} d\theta \xi^\mu (\theta) = 0 .
\eea
Then the $\delta$-function splits into factors 
corresponding to the different modes;
\bea
 \D(X^\mu (\theta) - f^\mu ( t(\theta), \s (\theta) ) )
	& = & \D ( x^\mu + \xi^\mu (\theta) 
	- f^\mu (t,\s) -\del_a f^\mu (t,\s) \z^a(\theta)- \cdots ) \nn \\
	& = & \D(x^\mu - f^\mu (t , \s))
	      \D(\xi^\mu (\theta) - \del_a f^\mu ( t, \s) \z^a (\theta)
- \cdots ) \nn \\
& = & \int D\nu_{0\mu} 
\exp \left[ i \int d\theta \nu_{0 \mu}(x^\mu - f^\mu (t , \s)) \right] \cdot
\nn \\
& & \int D\nu_{\mu}
\exp \left[ i \int d\theta \nu_{\mu}(\theta) (\xi^\mu (\theta) 
- \del_a f^\mu ( t, \s) \z^a (\theta)- \cdots ) \right]. \nn
\eea
Note that the fluctuation of $X^\mu$ 
comes from the change of the place 
where the open string is connected to the D1-brane,
but not from the change of the form of the function $f$, that is, 
the change of the form of the D1-brane.

 On the other hand,
we transform the fields $\xi^\mu$ to $\rho^a$
for which the determinant of the metric is $-1$.
\footnote
{ We can take $\rho^a$ to be the orthonormal coordinates \cite{kk}, 
but we do not so here for simplicity.}
\bea
 \xi^\mu (\s^{\bar{a}} ) & = & \hat{e}^{\mu}_A \rho^A (\s^{\bar{a}} ) , \\
x^A & = & \hat{e}^A_\mu x^\mu , \nn \\
\hat{e}^\mu_a & = & N_a^b \del_b f^{\mu}(t,\s) \vc{1} a,b = 0,1 , 
 \\
\hat{e}^\mu_\alpha & = & N_\alpha^\beta \mb{e}^\mu_\beta , \\
 & &  \hat{e}^\mu_a \hat{e}_{\mu \alpha} = 0 , \nn \\
 & &  \hat{e}^\mu_\alpha \hat{e}_{\mu \beta} = \delta_{\alpha \beta} . \nn
\eea
where $A$, $a$ and $\alpha$ denote all space-time directions, 
tangent and orthogonal directions to the D1-brane, respectively. 
Normalization factors are
\bea
N^A_B & = & \mb{diag}(N_0,N_1,1,\cdots ,1), \nn \\
 N_0 & = & \frac{1}{ \sqrt{-h_{00}}} , \vc{1} N_1  =  \sqrt{ \frac{h_{00}}{h}},
 \vc{1} h = \det (h_{ab}). 
\eea
In the same time
new tensors are defined as
\bea
\bar{h}_{AB} & = & 2\pi \alpha^\prime \hat{e}^\mu_A \hat{e}_{\mu B}
 =  \mb{diag}(\bar{h}_{ab},\delta_{\alpha \beta}) , \nn \\ 
& &  \bar{h}_{ab} = \hat{e}^\mu_a \hat{e}_{\mu b}  =  
\left[
\ba{cc}
N^2_0 h_{00} & N_0 N_1 h_{01}  \nn \\
N_0 N_1 h_{01} & N^2_1 h_{11} 
\ea
\right] , \nn
\\
& & \det (\bar{h}_{ab})  =  -1 , \\
\bar{F}_{AB} & = & 2\pi \alpha^\prime \hat{e}^\mu_A \hat{e}^\nu_B F_{\mu \nu},
\vc{1} \bar{F}_{ab}  \neq 0 ,\nn \\
\bar{u}_{AB} & = & 
2\pi \alpha^\prime \hat{e}^\mu_A \hat{e}^\nu_B \del_\mu \del_\nu T ,
\vc{1} \bar{u}_{ab} \neq 0 , \nn \\
\hat{\nu}_A & = & \hat{e}^B_A \nu_B ,
\vc{1} \hat{\nu}_{0 A}  =  \hat{e}^B_A \nu_{0 B}. \nn
\eea
In the following discussion
the metrics in (\ref{nold}) and (\ref{niold}) 
change from $\eta$ to $\bar{h}$;
The Neumann functions with the new metric are denoted 
by $\neu^{AB}$ and $\neu^{-1}_{AB}$.
It is supposed that
$A_\mu$ is nonzero only for the component tangent to the D1-brane,
and $A$ and $T$ depend only on the coordinates tangent to the D1-brane.
In this case
$\bar{F}_{\alpha A}$ and $\bar{u}_{\alpha A} $ vanish and then
$\bar{u}$ can be written in the covariant form
\bea
 \bar{u}_{ab} = 
2\pi \ap N^c_a N^d_b \del_c f^\mu \del_d f^\nu \del_\mu \del_\nu T
= 2\pi \ap N^c_a N^d_b \nabla_c \nabla_d T .
\label{covu}
\eea
 
By use of them,
the action is rewritten in terms of $\rho$:
\bea
S & = & \frac{1}{4\pi \alpha^\prime} 
 \left[ 
\rho^A \cdot \neu^{-1}_{AB} \cdot  \rho^B
+ i \int_{\del \Sigma} d\theta \bar{F}_{AB}(x^A) \rho^A \del_\theta \rho^B 
+ \frac{1}{\E} \int_{\del \Sigma} \bar{u}_{AB}(x^A)\rho^A \rho^B 
+ \cdots
\right] , \nn \\
& + & \frac{1}{\E} \int_{\del \Sigma } d\theta a(x^A) ,
\eea
where $\cdots$ means higher derivative terms of $A_\mu$ and $T$,
and we ignore them here.
In this action,
regarding the terms from the tachyon field, $a$ and $\bar{u}$, as perturbation,
\footnote{
If $\bar{u}$ is included in the Green function,
the Partition function looks like that in \cite{Oku}.
}
the conditions imposed on $\rho^a$ and $\rho^\alpha$ are
\bea
\neu^{-1}_{ab} \cdot \rho^b(\theta ) 
+ i \bar{F}_{ab} \del_\theta \rho^b(\theta )
= 0 \\
\neu^{-1}_{\alpha \beta} \cdot \rho^\beta(\theta ) 
= 0 .
\label{one-dim bulk cd}
\eea
On the other hand,
the conditions from the $\delta$-function are given by
\bea
 \rho^a(\theta) & = & (N^{-1})^a_b
 \left[ \z^b(\theta) 
- \frac{1}{3!} K^\mu_{cd}K_{\mu lm}h^{bc}\z^d(\theta) \z^l(\theta) \z^m(\theta) + \cdots
\right], \\
\rho_\alpha (\theta) & = & \hat{e}_{\alpha \mu}
\left[ 
\frac{1}{2}\del_a \del_b f^\mu \z^a(\theta) \z^b(\theta) 
\right.
\nn \\
& & + 
\left.
\frac{1}{3!} \left\{ \del_a\del_b \del_c f^\mu 
- 3\Gamma^d_{ab} \del_d \del_c f^\mu \right\}
\z^a(\theta) \z^b(\theta) \z^c(\theta) + \cdots
\right] .
\eea
From (\ref{one-dim bulk cd})
 we can put $\rho^\alpha$ as zero.

\vskip6mm\noindent
{\bf The Green function}

Let think about the Green function for (\ref{one-dim bulk cd}).
The Green function satisfies the following condition;
\bea
\neu^{-1}_{AB} \cdot \bar{M}^{BC} 
+ i \bar{F}_{AB} \del_\theta \bar{M}^{BC}
= \delta^C_A \delta(\theta -  \theta^\prime) .
\eea
A solution of this equation is \cite{t1}
\bea
\bar{M}^{AB}(\theta , \theta^\prime)
= \frac{1}{\pi}\mathop{ \sum^\infty_{k=1}}\frac{1}{n}\mb{e}^{-\E n}
[ G^{-1AB}\cos n (\theta -  \theta^\prime)
+ i \Theta^{AB} \sin n (\theta -  \theta^\prime) ] ,
\label{g1}
\eea
where $G^{-1}$ and $\Theta$ are respectively 
symmetric and  antisymmetric tensors
defined as
\bea
\left( \frac{1}{\bar{h} - \bar{F}} \right)^{AB}
 = G^{-1AB} + \Theta^{AB}.
\eea
The Green function (\ref{g1}) has the short distance divergence such that
\bea
 \bar{M}^{AB}(\theta ,\theta)
& = & 
\frac{1}{\pi}\mathop{ \sum^\infty_{k=1}}\frac{1}{n}\mb{e}^{-\E n} G^{-1AB}
\nn \\
& = & 
 -\frac{1}{\pi} G^{-1AB}\ln \E + {\cal O }( \ln (1-\E)).
\label{g1d}
\eea
The first term in (\ref{g1d}) diverges as $\E$ closes to zero.

\section{Path Integral}
 
\renewcommand{\theequation}{2.\arabic{equation}}
\vm{4}
Let carry out the calculations of the path integrals.
At first, we integrate with respect to $\rho^a$.
Since the integration of $\rho^\alpha$ give 
only a factor independent of $A$, $T$ and those derivatives,
we do not deal with it.

The part of the exponent of the partition function including $\rho^a$  is
\bea
I_{\rho^a} = \int D\rho^a \exp \left\{ \right.
& - &  \frac{1}{4\pi \ap } [ \rho^a \cdot \neu^{-1}_{ab} \cdot \rho^b
+ i\bar{F}_{ab} \rho^a \cdot \del_\theta \rho^b 
+ 2 \pi \ap \del_\rho (F_{\mu \nu} ) 
\hat{e}^\rho_a \hat{e}^\mu_b \hat{e}^\nu_c 
\rho^a \cdot \rho^b \del_\theta x^c \nn \\
& + & \left. \frac{1}{\E} \bar{u}_{ab} \rho^a \cdot \rho^b
+ { \cal O }( \rho^3 ) ]  + i \hat{\nu}_a \cdot \rho^a\right\}.
\label{rhopart}
\eea
We treat $\bar{u}$ and  $\del F$ as perturbation. 
Integrating with respect to $\rho^a$,
we have
\bea
I_{\rho^a} \propto 
[\mb{Det}^\prime ( \neu^{-1}_{ab} +i\bar{F}_{ab}\del_\theta)]^{-1}
[1 - i \pi \ap \int d\theta \del_\nu (F_{\rho \mu}) 
\hat{e}^\nu_a  \hat{e}^\rho_b \hat{e}^\mu_c \bar{M}^{ab}(\theta, \theta)
\del_\theta x^c \nn \\
\left.
-\frac{1}{2\E} \int d\theta \mb{tr}[ \bar{u} \bar{M} (\theta, \theta)] \right]
\exp [ -\pi \ap \hat{\nu}_a \cdot \bar{M}^{ab} \cdot \hat{\nu}_b ].
\eea
Det is defined as the determinant with respect to modes of the fields and
$\mb{Det}^\prime$ stands for also the determinant 
excluding the contributions from zero-modes.
det and tr are defined as that with subscripts $a$ and $b$.

Let integrate with respect to $\hat{\nu}_a$.
The part of $Z$ including $\hat{\nu}_a$ is
\bea
I_{\hat{\nu_a}} = \int D\hat{\nu}_a \exp 
[ -\pi \ap \hat{\nu}_a \cdot \bar{M}^{ab} \cdot \hat{\nu}_b 
- i \int d\theta \hat{\nu}_a J^a],
\eea
where
\bea
 J^a = (N^{-1})^a_e \left( \z^e - \frac{1}{3!}K^\mu_{ab} K_{\mu cd} h^{ea}
\z^b \z^c \z^d + {\cal O}(\z^4) \right) .
\eea 
Then we obtain
\bea
I_{\hat{\nu_a}} \propto 
[\mb{Det}^\prime(\bar{M}^{ab})]^{-1}
\exp \left\{ - \frac{1}{4\pi \ap} J^a \cdot \bar{M}^{-1}_{ab} \cdot 
J^b \right\}.
\eea
The factor $[\mb{Det}^\prime(\bar{M}^{ab})]^{-1}$ is canceled by the factor 
$[\mb{Det}^\prime ( \neu^{-1}_{ab} +i\bar{F}_{ab}\del_\theta)]^{-1}$.

Finally we integrate over $\z^a$.
In order to identify the derivative expansion 
with the $\alpha^\prime$ expansion,
we rescale $\z$ as follows:
\bea
 \z^a & = &  N^a_b \sqrt{\ap} \bar{\z^b} , \nn \\
 D\z^a &  = &  \frac{1}{\ap}\sqrt{-h} D\bar{\z}^a .
\eea
The exponent including $\bar{\z^a}$ is obtained as
\bea
& - & \frac{1}{4\pi} \int d\theta d\theta^\prime
\left[
\bar{\z}^a(\theta) D_{ab}(\theta, \theta^\prime) \bar{\z}^b(\theta^\prime)
\right. \nn \\
&& -  \left. \frac{\ap}{3}N^{b^\prime}_b N^{c^\prime}_c N^{d^\prime}_d 
(N^{-1})^e_{e^\prime}
 K^\mu_{a^\prime b^\prime}K_{\mu c^\prime d^\prime}h^{e^\prime a^\prime}
(\bar{M}^{-1}(\theta, \theta^\prime))_{ea} 
\bar{\z}^b(\theta)\bar{\z}^c(\theta) \bar{\z}^d(\theta) 
\bar{\z}^a(\theta^\prime)
\right] \nn \\
& + & {\cal O}(\alpha^{\prime \frac{3}{2}}),
\eea
where
\bea
D_{ab}(\theta, \theta^\prime)
=
(\bar{M}^{-1}(\theta, \theta^\prime))_{ab}
 + i 2\pi \ap N_a^{a^\prime} N_b^{b^\prime}
\hat{\nu}^\alpha (\theta) \hat{e}_{\alpha \mu}
\del_{a^\prime} \del_{b^\prime} f^\mu \delta(\theta - \theta^\prime).
\eea
Since $\hat{\nu}^\alpha$ does not have the zero mode and
$ \bar{M}^{ab}(\theta, \theta)$ is independent of $\theta$,
$\int d\theta \hat{\nu}^\alpha(\theta)\bar{M}^{ab}(\theta, \theta)$ vanishes.
Then, $\mb{Det}^\prime (D) = \mb{Det}^\prime (\bar{M}^{-1}) 
+ {\cal O}(\alpha^{\prime 2})$.

By integration with respect to $\z$,
 regarding $\ap$ higher order terms as perturbation,
the partition function becomes
\bea
Z \propto & &\int Dx Dt D\s  \delta (x^\mu - f^\mu(t, \s))
\sqrt{-h}
[\mb{Det}^\prime(\bar{M}^{-1})]^{-\frac{1}{2}}
\exp \left( - \frac{1}{\E} \int d\theta T(x) \right) \nn \\
& &
\{ 1 
 -  i \pi \ap \int d\theta \del_\nu (F_{\rho \mu})
\hat{e}^\nu_a  \hat{e}^\rho_b \hat{e}^\mu_c \bar{M}^{ab}(\theta, \theta)
\del_\theta x^c \nn \\
& & -  \frac{1}{2\E} \int d\theta
\left[  \bar{u}_{ab} \bar{M}^{ba} (\theta, \theta) \right. \nn \\
& & +  \frac{\ap}{3}\pi [ 2N^{b^\prime}_b N^{c^\prime}_c
K^\mu_{a^\prime b^\prime}K_{\mu c^\prime d^\prime}
h^{d^\prime a^\prime} \bar{M}^{bc}(\theta, \theta)
+ 
N^{c^\prime}_c N^{d^\prime}_d
K^\mu_{a^\prime b^\prime}K_{\mu c^\prime d^\prime}
h^{a^\prime b^\prime} \bar{M}^{cd}(\theta, \theta) ]  \nn \\
& & +  \left. \left. {\cal O}(\alpha^{\prime \frac{3}{2}}) \right] \right\},
\label{f1}
\eea
where
\bea
[\mb{Det}^\prime(\bar{M}^{-1})]^{-\frac{1}{2}}
= [\mb{Det}^\prime(\bar{M})]^{\frac{1}{2}}
= \sqrt{\mb{det}(\bar{h}+\bar{F})} .
\label{BI factor}
\eea
In the calculation of (\ref{BI factor})
 we used the zeta-function regularization.

It is shown that the $\ap$ reading terms include divergences and
the partition function is the same form as that on the flat D-brane
when these divergent terms are canceled.

\vskip6mm\noindent
{\bf $\beta$ function of tachyon}

 We have shown that the $\ap$ leading terms
 include divergences.
Since the term proportional to $\del(F)$ vanish when $F$ 
 is constant,
the divergence included in the second term in (\ref{f1}) is ignored.
We renormalize the tachyon field to cancel
other two terms in (\ref{f1}).
The renormalized tachyon denoted by $T_R$ is defined in the $\ap$ order 
as follows:
\bea
T(x) = \E \left[ T_R \right. & + & \mb{tr}[ \bar{u} G^{-1}] \ln \E \nn \\
& + & \left. \frac{\ap}{3}(2 N^{b^\prime}_b N^{c^\prime}_c
K^\mu_{a^\prime b^\prime}K_{\mu c^\prime d^\prime}
h^{d^\prime a^\prime} G^{-1bc}
+ N^{c^\prime}_c N^{d^\prime}_d
K^\mu_{a^\prime b^\prime}K_{\mu c^\prime d^\prime}
h^{a^\prime b^\prime} G^{-1cd} ) \ln \E \right] . \nn \\
\label{te}
\eea
Substituting (\ref{covu}) to (\ref{te}),
the beta-function of a tachyon becomes
\bea
\beta_T  = 
 - \left[ T_R \right. & + & 2\pi \ap N_a^{a^\prime} N_b^{b^\prime}
G^{ -1 ab}\nabla_{a^\prime} \nabla_{b^\prime} T_R  \nn \\
& + & \left. \frac{\ap}{3}(2 N^{b^\prime}_b N^{c^\prime}_c
K^\mu_{a^\prime b^\prime}K_{\mu c^\prime d^\prime}
h^{d^\prime a^\prime} G^{-1bc}
+ N^{c^\prime}_c N^{d^\prime}_d
K^\mu_{a^\prime b^\prime}K_{\mu c^\prime d^\prime}
h^{a^\prime b^\prime} G^{-1cd} ) \right].  \nn \\
\label{betat}
\eea
The $\beta_T$is  different from that of the case of a flat brane 
by an inhomogeneous term.
Because of the term the tachyon satisfies the equation 
different from that of the flat brane 
under the condition that the beta function vanishes.
We see that 
the geometrical information of the brane configuration is 
included in the tachyon.

The effective action $S$ including the tachyon is related 
to the partition function $Z$
\cite{t1,sz}.
Coefficients of 
the divergent terms in (\ref{f1}),
the tachyon kinetic term and the term including the extrinsic curvatures,
are scheme dependent
and we can determine these coefficients from the claim that 
$\frac{\delta S}{\delta T} = 0 $
is equivalent to 
$\beta_T = 0$
\cite{t1,sz2}.
In general
this calculation is complicated.
Exceptionally in the case
where the last term in (\ref{betat}), 
the normalization factor $N^a_b$,
the induced metric $\bar{h}_{ab}$ and the field strength $\bar{F}_{ab}$ are 
constant (for example when the D1-brane forms a tube),
we can determine these coefficients as in the flat case
\footnote{
Alternately we define the new field
\be
 \tilde{T} = T + 
\frac{\ap}{3}(2 N^{b^\prime}_b N^{c^\prime}_c
K^\mu_{a^\prime b^\prime}K_{\mu c^\prime d^\prime}
h^{d^\prime a^\prime} G^{-1bc}
+ N^{c^\prime}_c N^{d^\prime}_d
K^\mu_{a^\prime b^\prime}K_{\mu c^\prime d^\prime}
h^{a^\prime b^\prime} G^{-1cd} ).
\ee
Then $\beta_T = 0 $  means that $\tilde{T}$ satisfies 
the ordinary equation of motion of the tachyon on a flat brane and 
we can obtain the effective action as that on the flat brane. 
}.

\vskip6mm\noindent
{\bf Acknowledgement}

\vskip2mm
We would like to thank Dr. M. Naka for encouragement.
We also grateful to Dr. Y. Satoh and Dr. K. Mohri 
for helpful discussions on several points in this paper. 
And we thank to Prof. T. Kobayashi for valuable comments.

\newpage


\end{document}